\documentclass[a4paper]{article}
\usepackage{amsmath}
\usepackage{breqn}
\usepackage{amsmath, graphicx}
\usepackage{comment}
\usepackage{cite}
\usepackage{enumerate}
\usepackage{array}
\usepackage{breqn}
\usepackage{amssymb}

\usepackage{multirow}
\usepackage{arydshln}
\usepackage{xcolor}
\usepackage{INTERSPEECH2019}
\ninept
\title{Robust Raw Waveform Speech Recognition Using \\
Relevance Weighted Representations}
\name{Purvi Agrawal and Sriram Ganapathy
\thanks{This work was partly funded by grants from the Department of Science and Technology project DST0 (ECR01341), Govt. of India,
and Indian Institute of Science.
}
}
\address{
  Learning and Extraction of Acoustic Patterns (LEAP) Lab, Dept. of Electrical Engg., \\ Indian Institute of Science, Bengaluru-560012, India.
  }
\email{(purvia, sriramg)@iisc.ac.in}
\begin{document}

\maketitle
\begin{abstract}
Speech recognition in noisy and channel distorted scenarios is often challenging as the current acoustic modeling schemes are not adaptive to the changes in the signal distribution in the presence of noise. In this work, we develop a novel acoustic modeling framework for noise robust speech recognition based on relevance weighting mechanism. The relevance weighting is achieved using a sub-network approach that performs feature selection. A relevance sub-network is applied on the output of first layer of a convolutional network model operating on raw speech signals while a second relevance sub-network is applied on the second convolutional layer output. The relevance weights for the first layer correspond to an acoustic filterbank selection while the relevance weights in the second layer perform modulation filter selection. The model is trained for a speech recognition task on noisy and reverberant speech.
The speech recognition experiments on multiple datasets (Aurora-4, CHiME-3, VOiCES) reveal that the incorporation of relevance weighting in the neural network architecture improves the speech recognition word error rates significantly (average relative improvements of $10$\% over the baseline systems).
\end{abstract}
\noindent\textbf{Index Terms}: Raw speech waveform, relevance weighting, cosine-modulated Gaussian filterbank, speech recognition.

\section{Introduction}
\label{sec:intro}
The broad set of methods that enable the learning of meaningful representations for a given data are referred to as representation learning methods. This can be unsupervised like principal components or supervised like linear discriminant analysis. With the growing interest in deep learning, representation learning using deep neural networks has been actively pursued. While a lot of success has been reported for text and other domains (for example, using word2vec models \cite{mikolov2013efficient}), representation learning for speech is still challenging. This paper explores representation learning for speech using a novel modeling approach.

In the past, the main direction pursued  has been to learn filterbank parameters \cite{doss2013, sainath2013, tuske2014acoustic} from raw waveforms. The objective can be either detection or classification \cite{sainath2013, hoshen2015speech, sainath2015cldnn}. 
Some of the efforts also attempt unsupervised learning of filterbank, eg. Sailor et. al \cite{sailor2016filterbank} uses  restricted Boltzmann machine while Agrawal et. al. \cite{agrawal2019unsupervised} uses variational autoencoders. The wav2vec method by Schneider et. al. in \cite{schneider2019wav2vec} explores unsupervised pre-training for speech recognition by learning representations of raw audio.
There has been some attempts to explore interpretability of acoustic filterbank recently, for eg. SincNet filterbank \cite{ravanelli2018interpretable, pascual2019pase}.  However, compared to vector representations of text which have shown to embed meaningful semantic properties, the interpretability of speech representations from these approaches has often been limited. Further, most of the state-of-the-art systems continue to use mel filterbank \cite{mfccdavis} features.

The approach of modulation filter learning  (modulation filters process the time-frequency representation and perform filtering along time (rate) and frequency (scale) dimensions) using the linear discriminant analysis (LDA) has been explored to learn the
temporal modulation filters in a supervised manner \cite{vuurenLda1997, hung2006optimization}. There have also been attempts to learn modulation filters in an unsupervised manner \cite{purvi2017MF1D, sailor2016unsupervised,  agrawal2019jstsp, purvi2019skipConn}. 

In this paper, we propose a relevance weighting mechanism that allows the interpretability of the learned representations in the forward propagation itself. 
The relevance weighting scheme is popular in text domain in applications such as document search, where  a static relevance weight is attached to each document, based on the search term feature \cite{robertson1976relevance, robertson1997relevance}. A similar application of visual attention in the image domain uses spatial weighting to weigh different parts of the image 
\cite{xu2015show}. A related work is the deep mixture of experts (MoE) model that trains multiple expert networks, each of which specializes in a different part of the input space and a gating network decides which expert to use for each input region \cite{eigen2013learning}. In this work, the relevance weighting on the learned representations is achieved using a sub-network. 

We propose a  speech representation learning method using a two-step relevance weighting approach. The first step performs relevance weighting on the output of the first convolutional layer that learns acoustic filterbank from the raw waveform. The acoustic filters are parametric cosine-modulated Gaussian filters whose parameters are learned within the acoustic model \cite{agrawal2019unsupervised}. 
The output is fed to the relevance sub-network to obtain the relevance weights for the filterbank outputs. The weighted filterbank representation is used as input to the second convolutional layer which is interpreted as a modulation filtering step. 
The kernels of the second convolutional layer are 2-D spectro-temporal modulation filters and the filtered representations are weighted using another relevance sub-network.
The rest of the architecture performs the task of acoustic modeling for automatic speech recognition (ASR). 
All the model parameters 
are learned in a supervised fashion.  
The ASR experiments are conducted on Aurora-4 (additive noise with channel artifact) \cite{hirsch2000aurora}, CHiME-3 (additive noise with reverberation) \cite{barker2015chime3} and VOiCES (additive noise with reverberation) \cite{nandwana2019voices} databases. The experiments show that the learned representations from the proposed  framework 
provides considerable improvements in ASR results over the baseline methods.

The rest of the paper is organized as follows. 
Section \ref{sec:rep_learning} describes the proposed two-step representation learning approach using relevance weighting. 
Section~\ref{sec:experiments} describes the ASR experiments with the various front-ends 
followed by a summary.



\begin{center}
    \begin{figure}[t]
        \centering
        \includegraphics[trim={0.1in 12in 8.1in 0in}, clip, width=\linewidth]{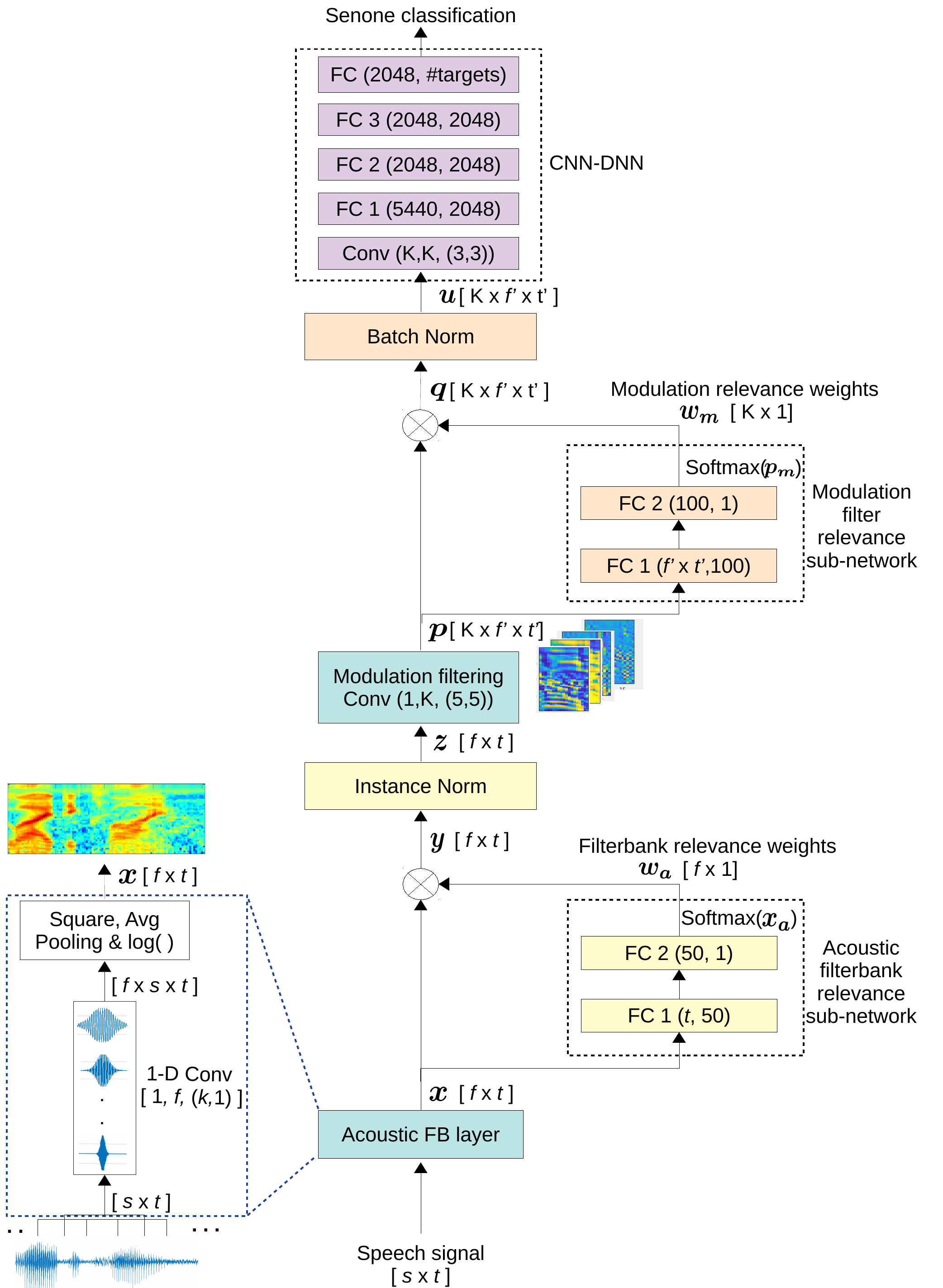} \vspace{-0.5cm}
        \caption{Block diagram of the proposed representation learning from raw waveform using relevance weighting approach. Here, FC represents fully connected layer. }
        \label{fig:block_diag}
        \vspace{-0.3cm}
    \end{figure}
\end{center}
\vspace{-0.9cm}
\section{Relevance Based Representation Learning}{\label{sec:rep_learning}}
The block schematic of the proposed relevance weighting based two-step representation learning model is shown in Figure~\ref{fig:block_diag}.

\subsection{Step-1: Acoustic Filterbank representation} 
 The input to the neural network are raw samples windowed into $s$ samples per frame with a contextual window of $t$ frames. Each block of $s$ samples is referred to as a frame. This matrix of size $s \times 1$ raw audio samples are processed with a 1-D convolution  using $f$ kernels ($f$ denotes the number of sub-bands in filterbank decomposition) each of size $k$.  The kernels are modeled as cosine-modulated Gaussian function \cite{agrawal2019unsupervised},
\begin{equation}
    {{g}}_i (n) = \cos{2\pi\mu_i n} \times \exp{(-{n^2}\mu_i^2/{2})}
\end{equation}
where ${{g}}_i (n)$ is the $i$-th kernel ($i=1,..,f$)  at time $n$, $\mu_i$ is the center frequency of the $i$th filter (in frequency domain).
The parametric approach to filterbank (FB) learning generates filters with a smooth frequency response.
The mean parameters are updated in a supervised manner for each dataset.
The convolution with the cosine-modulated Gaussian filters generates $f$ feature maps. These outputs are squared, average pooled within each frame and log transformed. This generates $\boldsymbol{x}$ as $f$ dimensional features for each of the $t$ contextual frames, as shown in Figure \ref{fig:block_diag}. The $\boldsymbol{x}$ can be interpreted as the ``learned'' time-frequency representation (spectrogram). We refer to the first layer as the acoustic filterbank (FB) layer.
\subsection{Acoustic FB relevance weighting} {\label{subsec:self_relevance weighting}}
The relevance weighting paradigm for acoustic FB layer is implemented using a relevance sub-network fed with the $f \times t$ time-frequency representation $\boldsymbol{x}$. A two layer deep neural network (DNN) with a softmax output generates acoustic FB relevance weights $\boldsymbol{w}_a$ as $f$ dimensional vector with weights corresponding to each sub-band filter. Let the output from the relevance sub-network be denoted as $\boldsymbol{x}_a$, then the relevance weights $\boldsymbol{w}_a$ are generated using the softmax function as,
\begin{equation}
    w_a^i = \frac{e^{x_a^i}}{\sum_j e^{x_a^j}}; ~~\text{where}~ i = 1, 2, ..., f. 
\end{equation}
These weights $\boldsymbol{w}_a$ are multiplied element-wise with each frame of $\boldsymbol{x}$ to obtain weighted filterbank representation $\boldsymbol{y}$. The relevance weights in the proposed framework are different from typical relevance weights used in text search problem \cite{robertson1997relevance} as well as the attention mechanism~\cite{zhang2017attention}. In proposed framework, relevance weighting is applied on the representation as soft feature selection weights without performing a linear combination. We also smooth the first layer outputs ($\boldsymbol{y}$) using instance norm~\cite{rumelhart1986learning, ulyanov2016instance}. Let $y_{j,i}$ denote the relevance weighted filterbank output for frame $j$ ($j=1,..,t$) of sub-band $i$ ($i=1,..,f$). The soft weighted output $z_{j,i}$ is given as,
\begin{equation}\label{eq:soft}
    z_{j,i} = \frac {y_{j,i} - m_{i}} {\sqrt{\sigma ^2 _i + c}} 
\end{equation}
where $m_i$ is the sample mean of $y_{j,i}$ computed over $j$ and $\sigma _i$ is the sample std. dev. of $y_{j,i}$ computed over $j$. The constant $c$ is $1e-4$. 
The output of relevance weighting ($\boldsymbol{z}$) is propagated to the subsequent layers for the acoustic modeling. 

In our experiments, we use $t=101$ whose center frame is the senone target for the acoustic model. We also use $f=80$ sub-bands and acoustic filter length $k=129$. This value of $k$ corresponds to $8$ ms in time for a $16$ kHz sampled signal which has been found to be sufficient to capture temporal variations of speech signal \cite{lewicki2002efficient}. The value of $s$ is $400$ corresponding to $25$ ms window length and the frames are shifted every $10$ms. Thus, the input to the acoustic filter bank layer with $t=101$ contains about $1$ sec.  of audio segment. In our experiments, we also find  that after the normalization layer, the number of frames $t$ can be pruned to the center $21$ frames for the acoustic model training without loss in performance. This has significant computational benefits and the pruning is performed to keep only the $21$ frames around the center frame ($200$ ms of context). 

\begin{figure}[t]
    \centering
    \vspace{-0.1cm}
    \includegraphics[trim={1cm 1.85in 0cm 1cm}, clip, scale=0.33]{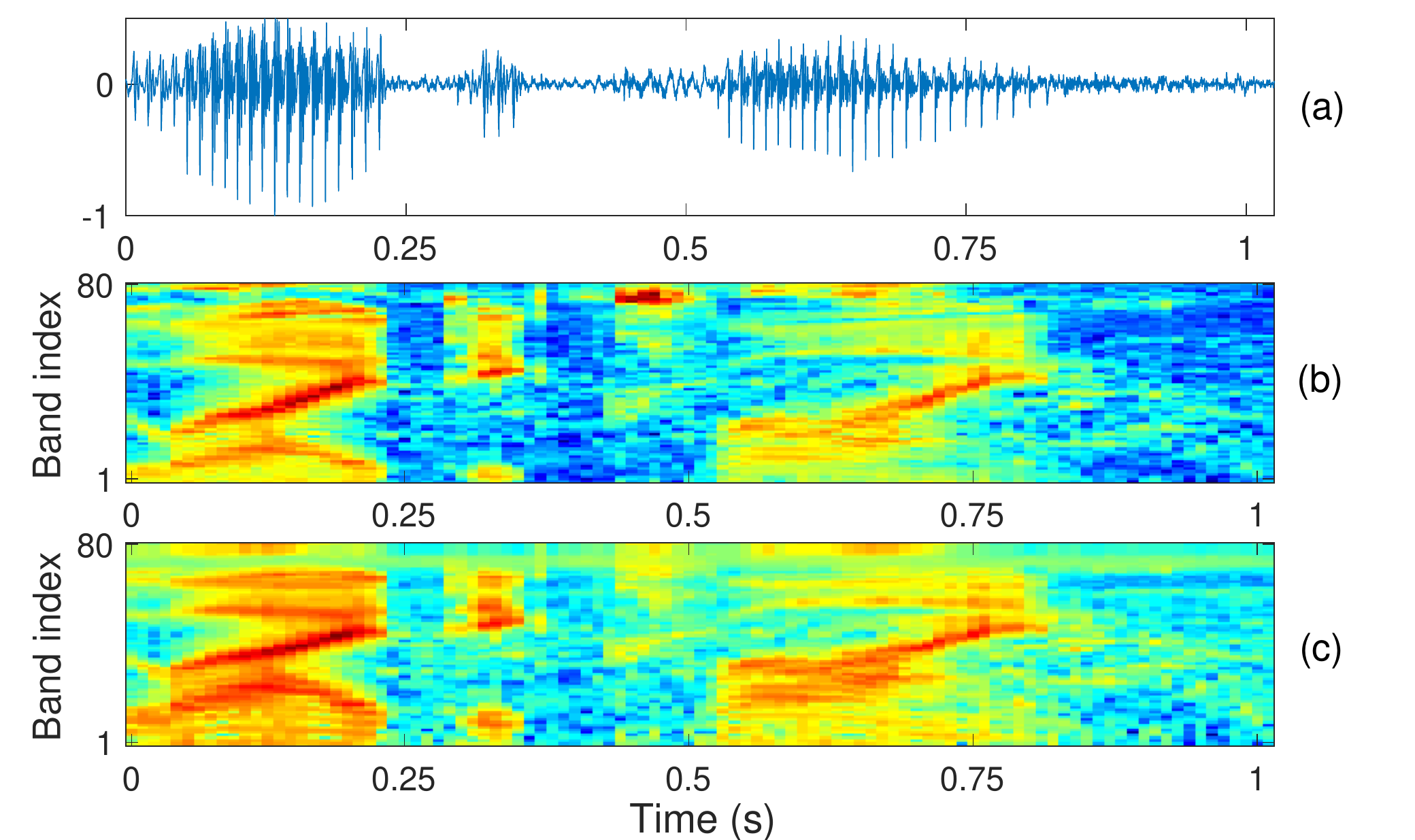}
    \vspace{-0.2cm}
    \caption{(a) Speech signal from Aurora-4 dataset with airport noise, (b) mel spectrogram representation (c) acoustic FB representation with soft relevance weighting ($\boldsymbol{z}$ in Figure \ref{fig:block_diag}).}
    \label{fig:sig_spec} 
    \vspace{-0.4cm}
\end{figure}

The soft relevance weighted time-frequency representation $\boldsymbol{z}$ obtained from the proposed approach is shown in Figure \ref{fig:sig_spec}(c) for an utterance with airport noise from Aurora-4 dataset (the waveform is plotted in Figure \ref{fig:sig_spec}(a)). The corresponding mel spectrogram (without relevance weighting) is plotted in Figure \ref{fig:sig_spec}(b). It can be observed that, in the learned filterbank representation (Figure \ref{fig:sig_spec}(c)), the formant frequencies appear to be shifted upwards because of the increased number of filters in the lower frequency region. Also, the relevance weighting modifies the representations propagated to the higher layers.

\subsection{Step-2: Relevance Weighting of Modulation Filtered Representation}
The representation $\boldsymbol{z}$ from acoustic filterbank layer is fed to the second convolutional layer which is interpreted as modulation filtering layer (shown in Figure \ref{fig:block_diag}). The kernels of this convolutional layer are interpreted as 2-D spectro-temporal modulation filters, learning the rate-scale characteristics from the data. This step is partly inspired by the neuro-physiological evidences of multi-stream feature framework for ASR \cite{mesgarani2006discrimination, nemala2013multistream}. The modulation filtering layer generates $K$ parallel streams, corresponding to $K$ modulation filters $\boldsymbol{w}_K$. The modulation filtered representations $\boldsymbol{p}$ are max-pooled with window of $3 \times 1$, leading to feature maps of size $f' \times t'$. These are weighted using a second relevance weighting sub-network (referred to as the modulation filter relevance sub-network in Figure \ref{fig:block_diag}). Let $\boldsymbol{p}_m$ denote the $K$-dimensional output of modulation filter relevance sub-network. The softmax function is applied on the output to generate modulation relevance weights $\boldsymbol{w}_m$ over $K$ modulation filters,
\vspace{-0.1cm}
 \begin{equation}
     w_m^i = \frac{e^{p_m^i}}{\sum_j e^{p_m^j}}; ~\text{where}~ i = 1, 2, ..., K. 
 \end{equation}
The weights are multiplied with the representation $\boldsymbol{p}$ to obtain weighted representation $\boldsymbol{q}$. This weighting is interpreted as the selection of different modulation filtered representations (with different rate-scale characteristics). The resultant weighted representation $\boldsymbol{q}$ is fed to the batch normalization layer \cite{batch2015norm}. The training data statistics of batch norm, including affine parameters, 
are used in the test phase. 
The value of the normalization factor $c$ in denominator for batch norm is chosen to be $10^{-4}$ empirically. We use the value of $K=40$ in the work. Following the acoustic filterbank layer and the modulation filtering layer (including the relevance sub-networks), the acoustic model consists of series of CNN and DNN layers. The configuration details are given in Figure~\ref{fig:block_diag}. 

The proposed two stage processing is loosely modeled based on our understanding of the human auditory system, where the cochlea performs acoustic frequency analysis while early cortical processing performs modulation filtering \cite{mesgarani2006discrimination}. The relevance weighting mechanism attempts to model the feature selection/weighting inherently present in the auditory system (based on the relative importance of the representation for the downstream task). 
\section{Experiments and Results}{\label{sec:experiments}}
The speech recognition system is trained using PyTorch \cite{paszke2017pytorch} while the Kaldi toolkit~\cite{povey2011kaldi} is used for decoding and language modeling. The ASR is built on three datasets, Aurora-4, CHiME-3 and VOiCES respectively. The models are discriminatively trained using the training data with cross entropy loss and Adam optimizer \cite{kingma2014adam}. A hidden Markov model - Gaussian mixture model (HMM-GMM) system is used to generate the senone alignments for training the CNN-DNN based model.
The ASR results are reported with a tri-gram language model and the best language model weight is obtained from the development set. 

For each dataset, we compare the ASR performance of the proposed approach of learning acoustic representation from raw waveform with acoustic FB (A) with relevance weighting (A-R) and modulation FB (M) with relevance weighting (M-R) denoted as (A-R,M-R),
with the model having only the acoustic FB relevance weighting (A-R), traditional mel filterbank energy (MFB) features, and power normalized filterbank energy (PFB) features \cite{kim2012pncc}. For CHiME-3 dataset, we also compare with RASTA features that perform modulation filtering (RAS) \cite{hermanskyb}, and mean Hilbert envelope (MHE) features \cite{mhec2015}. All the baseline features are processed with cepstral mean and variance normalization (CMVN) on a $1$ sec. running window. The neural network architecture shown in Figure \ref{fig:block_diag} (except for the acoustic filterbank learning layer, the acoustic FB relevance  sub-network and modulation filter relevance  sub-network) is used for all the baseline features. 
\begin{center}   
     \begin{table}[t]
            \centering
            \begin{center}
            \caption{Word error rate (\%) in Aurora-4 database for multi-condition training with various feature extraction schemes.}
            \label{tab:multiData_aurora4}
            \vspace{-0.3cm}
            \resizebox{\columnwidth}{!}{
            \begin{tabular}{l|c|c|c|c|c|c|c|c}
            \hline
            Cond & MFB & PFB  & Sinc & A & MFB-R & A-R & S-R,M-R & A-R,M-R\\ \hline
            \multicolumn{9}{c}{A. Clean with same Mic} \\ \hline
            Clean & 4.2 & 4.0 & 4.0 & 4.1 & 4.0 & \textbf{3.6} & 3.8 & \textbf{3.6}\\ \hline
            \multicolumn{9}{c}{B: Noisy with same Mic} \\ \hline
            Airport & 6.8 & 7.1 & 6.9 & 6.4 & 7.0 & {6.0}  & 6.3 &  \textbf{5.9} \\
            Babble & 6.6 & 7.4 & 6.7 & 6.3 & 6.8 & \textbf{6.1} & 6.2  & \textbf{6.1} \\
            Car & {4.0} & 4.5 & 4.0 & {4.0} & 4.2 & {4.0}  & \textbf{3.9} & \textbf{3.9} \\
            Rest. & 9.4 & 9.6 & 9.4 & 8.5 & 9.4 & {7.7} & 8.4  & \textbf{6.8} \\
            Street & 8.1 & 8.1 & 8.4 & 7.8 & 8.0 & {7.1} & 7.5 & \textbf{6.9} \\
            Train & 8.4 & 8.6 & 8.3 & 7.9 & 8.6 & {7.3}  & 7.4 & \textbf{7.2} \\\hdashline
            Avg. & 7.2 & 7.5 & 7.3 & 6.8 & 7.3 & {6.4}  & 6.6 & \textbf{6.1}\\
             \hline
            \multicolumn{9}{c}{C: Clean with diff. Mic}\\ \hline
		    Clean & 7.2 & 7.3 & 7.3 & 7.3 & {7.1} & {8.1} & 6.8 & \textbf{6.0} \\ \hline
            \multicolumn{9}{c}{D: Noisy with diff. Mic}  \\ \hline
            Airport & 16.3 & 18.0 & 16.2 & 17.3 & 16.6 & {15.4} & \textbf{13.9} & {14.1} \\
            Babble & 16.7 & 18.9 & 17.6 & 17.4 & 16.7 & {16.0} & 16.0 & \textbf{15.4}\\
            Car & {8.6} & 11.2 & 9.0 & 9.0 & 9.0 & {9.4} & 7.9 & \textbf{7.7}\\
            Rest. & 18.8 & 21.0 & 19.0 & 18.2 & 18.5 & \textbf{16.9} & 19.2 & 18.6 \\
            Street & 17.3 & 19.5 & 17.3 & 17.8 & 17.5 & {16.9}  & \textbf{16.6} & {16.8}\\
            Train\ & 17.6 & 18.8 & 18.1 & 17.8 & 18.1 & \textbf{16.2} & 16.6 & \textbf{16.2} \\ \hdashline
            Avg. & 15.9 & 17.9 & 16.2 & 16.2 & 16.1 & {15.1} & 15.0  & \textbf{14.8} \\
             \hline
            \multicolumn{9}{c}{Avg. of all conditions}  \\ \hline
            Avg. & 10.7 & 11.7 & 10.8 & 10.7 & 10.8 & {10.0} & 10.0 & \textbf{9.6}\\ \hline
            \end{tabular}
            }
          \end{center}
          \vspace{-0.2cm}
      \end{table}
  \end{center}
\vspace{-0.7cm}
\subsection{Aurora-4 ASR}
This database consists of continuous read speech recordings of $5000$ words corpus, recorded under clean and noisy conditions (street, train, car, babble, restaurant, and airport) at $10-20$ dB SNR. The training data has $7138$ multi condition recordings ($84$ speakers) with total $15$ hours of training data. The validation data has $1206$ recordings for multi condition setup. The test data has $330$ recordings ($8$ speakers) for each of the $14$ clean and noise conditions. The test data are classified into group A - clean data, B - noisy data, C - clean data with channel distortion, and D - noisy data with channel distortion.

The ASR  performance on the Aurora-4 dataset is shown in Table \ref{tab:multiData_aurora4} for each of the $14$ test conditions. We also compare the ASR performance with the acoustic filterbank representation (A) without relevance weighting. In addition, we also experiment with the application of the relevance weighting over pre-trained mel filterbank features (MFB-R).

As seen in the results, most of the noise robust front-ends do not improve over the baseline mel filterbank (MFB) performance. The raw waveform acoustic FB performs similar to MFB baseline features on average while performing better than the baseline for Cond. A and B. The MFB-R features, which denote the application of the acoustic FB relevance weighting over mel filterbank features, also doesn't improve over baseline MFB features. The features with acoustic filterbank learning + relevance weighting (A-R) improves over the raw (A) features with average relative improvements of $6$\%. The proposed (A-R,M-R) representation learning (two-stage relevance weighting) provides  considerable improvements in ASR performance over the baseline system with average relative improvements of $11$\% over the baseline MFB features. Furthermore, the improvements in ASR performance are consistently seen across all the noisy test conditions. 

We also compare with the SincNet method \cite{ravanelli2018interpretable} where our cosine modulated Gaussian filterbank is replaced with the sinc filterbank \footnote{https://github.com/mravanelli/SincNet/} as kernels in first convolutional layer (acoustic FB layer in Fig. \ref{fig:block_diag}). The ASR system with sinc FB (Sinc) is trained jointly without any relevance weighting, and with 2-stage relevance weighting (S-R,M-R) keeping rest of the architecture same as shown in Fig. \ref{fig:block_diag}. From results in Table \ref{tab:multiData_aurora4}, it can be observed that the parametric sinc FB (without weighting) performs similar to MFB and our acoustic FB features (A). The relevance weighting over sinc FB (S-R,M-R) improves over the baseline MFB with average relative improvements of $6$\%.

\begin{table}[t!]
\begin{center}
\caption{Word error rate (\%) in CHiME-3 Challenge database for multi-condition training (real+simulated).}
\label{tab:Chime3Results}
\vspace{-0.2cm}
    \resizebox{\columnwidth}{!}{
	\begin{tabular}{l|c|c|c|c|c|c}
	\hline
		Test Cond & MFB & PFB & {RAS} & {MHE} & A-R & A-R,M-R \\ \hline
		Sim\_dev & 12.9 & 13.3 & 14.7 & 13.0 & {12.5} & \textbf{12.0}\\ 
		Real\_dev & 9.9 & 10.7 & 11.4 & 10.2 & {9.9}  & \textbf{9.6}\\ \hdashline
		Avg. & 11.4 & 12.0 & 13.0 & 11.6  & {11.2}  & \textbf{10.8}\\ \hline \hline

		Sim\_eval & 19.8 & 19.4 & 22.7 &  19.7 & {19.2} &  \textbf{18.5}\\ 
		Real\_eval & 18.3 & 19.2 & 20.5 & 18.5  & {17.3}  & \textbf{16.6}\\\hdashline
		Avg. & 19.1 & 19.3 & 21.6 &  19.1 & {18.2}  & \textbf{17.5}\\ \hline 

	\end{tabular}
	}
\end{center}
\vspace{-0.4cm}
\end{table}
\begin{center}
    \begin{table}[t!]
        \centering
        \caption{WER (\%) for cross-domain ASR experiments.
        }
        \label{tab:cross_domain_ASR}
        \vspace{-0.2cm}
        \resizebox{6.4cm}{0.75cm}{
        \begin{tabular}{l|c|c}
        \hline
         \multirow{2}{*}{\textbf{Filters Learned on }} & \multicolumn{2}{c}{\textbf{ASR Trained and Tested on}} \\ \cline{2-3}
          & Aurora-4 & CHiME-3   \\\hline
         Aurora-4 & 9.6 & 14.3 \\
         CHiME-3 & 9.7 & 14.2  \\\hline
        \end{tabular}
        }
        \vspace{-0.4cm}
    \end{table}
\end{center}
\vspace{-0.99cm}
\subsection{{CHiME-3 ASR}}
The CHiME-3 corpus for ASR contains multi-microphone tablet device recordings from everyday environments, released as a part of 3rd CHiME challenge \cite{barker2015chime3}. Four varied environments are present - cafe (CAF), street junction (STR), public transport (BUS) and pedestrian area (PED). For each environment, two types of noisy speech data are present - real and simulated. The real data consists of $6$-channel recordings of sentences from the WSJ$0$ corpus spoken in the environments listed above. The simulated data was constructed by artificially mixing clean utterances with environment noises. The training data has $1600$ (real) noisy recordings and $7138$ simulated noisy utterances, constituting a total of $18$ hours of training data.
We use the beamformed audio in our ASR training and testing. The development (dev) and evaluation (eval) data consists of $410$ and $330$ utterances respectively. For each set, the sentences are read by four different talkers in the four CHiME-3 environments. This results in $1640$ ($410 \times 4$) and $1320$ ($330 \times 4$) real development and evaluation utterances. 

The results for the CHiME-3 dataset are reported in Table \ref{tab:Chime3Results}. The initial  approach of raw waveform filter learning with acoustic FB relevance weighting improves over the baseline system as well as the other noise robust front-ends considered here. The proposed approach of 2-stage relevance weighting over learned acoustic and modulation representations provides significant improvements over baseline features. On the average, the proposed approach provides relative improvements of $10$\% over MFB features in the eval set. 


\subsection{Representation transfer across tasks}
In a subsequent analysis, we perform a cross-domain ASR experiment, i.e., we use the acoustic filterbank learned from one of the datasets (either Aurora-4 or CHiME-3 challenge)  to train/test ASR on the other dataset. The results of these cross-domain filter learning experiments are reported in Table~\ref{tab:cross_domain_ASR}.
The rows in the table show the database used to learn the acoustic FB and the columns show the dataset used to train and test the ASR (all other  layers in Figure~\ref{fig:block_diag} are learned in the ASR task). The performance reported in this table are the average WER on each of the datasets. The results shown in Table~\ref{tab:cross_domain_ASR} illustrate that the filter learning process is relatively robust to the domain of the training data, suggesting that the proposed representation learning approach can be generalized for other ``matched'' tasks.

\subsection{VOiCES ASR}
\begin{center}
    \begin{figure}[t]
        \centering
        \includegraphics[trim={0 0 0 0.2cm}, clip, scale=0.33]{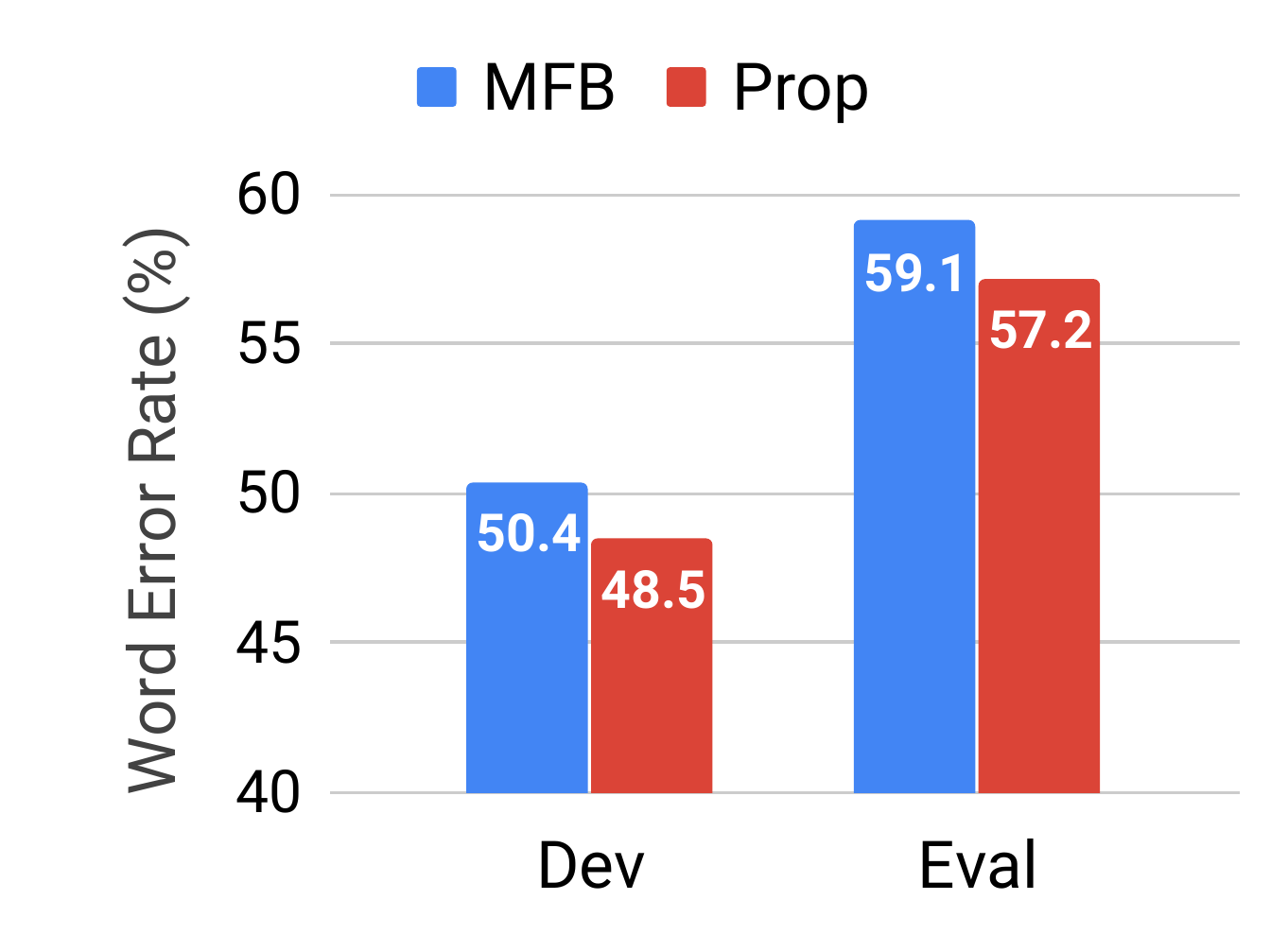}
        \vspace{-0.3cm}
        \caption{ASR performance in WER (\%) for VOiCES database.}
        \label{fig:voices_asr}
        \vspace{-0.5cm}
    \end{figure}
\end{center}

The Voices Obscured in Complex Environmental Settings (VOiCES) corpus is a creative commons speech dataset being used as part of VOiCES Challenge \cite{nandwana2019voices}.
The training data set of $80$ hours has $22,741$ utterances sampled at $16$kHz from $202$ speakers, with each utterance having $~12-15$s segments of read speech. 
We performed a 1-fold reverberation and noise augmentation of the data using Kaldi \cite{povey2011kaldi}.
The ASR development set consists of $20$ hours of  distant recordings from the $200$  VOiCES dev speakers. It contains recordings from $6$ microphones. The evaluation set  consists of $20$ hours of distant recordings from the $100$ VOiCES eval speakers and contains recordings from $10$ microphones. The ASR performance of VOiCES dataset with baseline MFB features and our proposed approach of 2-step relevance weighting is reported in Figure \ref{fig:voices_asr}. 
These results suggest that proposed model is also scalable to relatively larger ASR tasks with large vocabulary where consistent improvements can be obtained with the proposed approach.
\vspace{-0.9cm}

\vspace{-0.5cm} 
\section{Summary}{\label{sec:summary}}

The key contributions of the work are:
\begin{itemize}
\item Proposing a novel 2-stage relevance weighted representation learning neural architecture for speech modeling.  
\item The first stage weighs sub-bands of the learnt acoustic filterbank features from raw waveform; the second stage weighs the learnt modulation characteristics.
\item The weighting mechanism allows the feature selection and interpretability of the learnt representations in forward propagation itself.
\item Illustrating improved acoustic modeling using performance gains in word error rates for multiple ASR tasks.
\end{itemize}

\bibliographystyle{IEEEtran}

\bibliography{refer, refs}

\end{document}